\documentclass[aip,rsi,reprint,groupedaddress]{revtex4-1}
\usepackage[utf8x]{inputenc}
\usepackage{graphicx}
\usepackage{times}
\usepackage{wrapfig}
\usepackage{float}
\usepackage{graphicx,graphics,wrapfig,rotating}         
\usepackage{dcolumn}                                    
\usepackage{bm,fancybox}
\usepackage{amsmath,amssymb}                          
\usepackage{times,euscript,oldgerm}              %
\usepackage[english]{babel}                      %
\usepackage{psfrag}
\usepackage{hyperref}
\newcommand{\Yb}{$^{171}\rm{Yb}^{+}$}
\newcommand{\rep}{\nu_{rep}}
\newcommand{\dn}{|\Delta\nu_{M}|}

\begin{document}

\title{Beat note stabilization of mode-locked lasers \\ for quantum information processing}
\author{R. Islam, W. C. Campbell, T. Choi, S. M. Clark, S. Debnath, E. E. Edwards, B. Fields, D. Hayes, D. Hucul, \\
I. V. Inlek, K. G. Johnson, S. Korenblit, A. Lee, K. W. Lee, T. A. Manning, D. N. Matsukevich, J. Mizrahi, \\ 
Q. Quraishi, C. Senko, J. Smith, and C. Monroe}
\affiliation{Joint Quantum Institute, University of Maryland Department of Physics and \\
                    National Institute of Standards and Technology, College Park, MD  20742}
\date{\today}

\begin{abstract}
We stabilize a chosen radiofrequency beat note between two optical fields derived from the same mode-locked laser pulse train, in order to coherently manipulate quantum information. This scheme does not require access or active stabilization of the laser repetition rate. We implement and characterize this external lock, in the context of two-photon stimulated Raman transitions between the hyperfine ground states of trapped \Yb quantum bits.
\end{abstract}
\maketitle

\maketitle

The field of quantum information science holds great promise to revolutionize a wide range of applications, from computing and communication \cite{nielsen} to the simulation of complex quantum phenomena that cannot be modeled classically \cite{Feynman}. Quantum information is usually represented by distinct energy levels within quantum systems such as atoms or isolated solid state systems, with energy separations in the microwave or optical domain.  Such qubit states can typically be manipulated with external fields such as microwaves or optical fields whose frequency matches the qubit splitting, given appropriate couplings between the states.  Mode-locked lasers are versatile instruments for this purpose with their broadband optical spectra that feature both radiofrequency (rf) and microwave structure.  Such lasers have been used to control atomic \cite{HayesComb}, molecular \cite{Ye2007}, and solid-state quantum systems \cite{DeGreve2013}.  In this note, we describe a simple technique to phase-lock these optical sources in order to manipulate and control generic qubit systems \cite{Ladd2010}.

Mode-locked lasers can be used to produce a broadband optical frequency comb with overall bandwidth ranging from $10$~GHz -- $100$~THz with comb teeth that are spaced by the repetition rate of the laser, typically in the range $0.1-1$ GHz.  In order to bridge the frequency gap of a particular qubit, the qubit splitting is matched to be near an integral multiple of the repetition rate, and fine tuning is accomplished with additional optical modulators \cite{HayesComb}.  In order to retain long term coherence in the quantum system, the relevant optical frequency difference from the laser source must be stable.  This can be accomplished by directly locking the laser repetition rate by controlling the laser cavity spacing.  Here we describe a more useful technique to stabilize the beat note using acousto-optic modulators (AOMs) outside the laser cavity. This technique provides higher lock bandwidths without requiring invasive access to the laser cavity. 

\begin{figure}
\begin{center}
\includegraphics[width=0.2\textwidth]{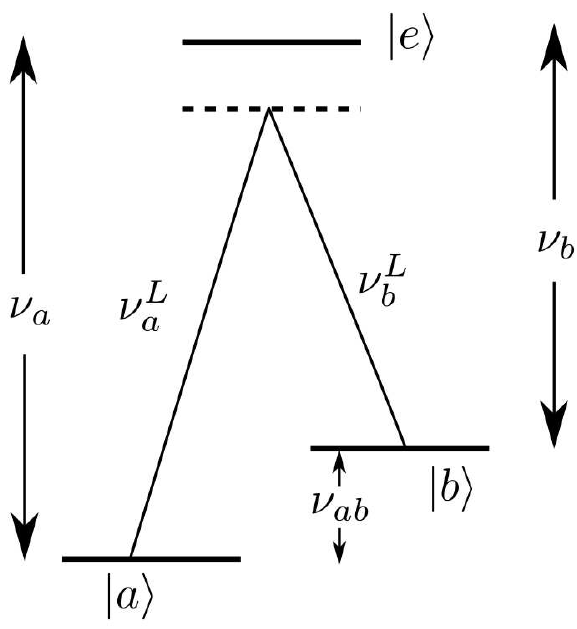}
\end{center}
\caption{Energy level diagram in a three level $\Lambda$ system.  Two laser beams with frequencies $\nu_{a}^{L}$ and $\nu_{b}^{L}$ off-resonantly couple the states $|a\rangle$ and $|b\rangle$ to the excited state $|e\rangle$. When the detunings $\nu_{a}-\nu_{a}^{L}$ and $\nu_{b}-\nu_{b}^{L}$ are larger than the strength of the couplings, the excited state may be adiabatically eliminated and the system behaves like an effective two level qubit with states $|a\rangle$ and $|b\rangle$. } 
\label{fig:level_diagram}
\end{figure}

We consider the control of a generic three level `$\Lambda-$' quantum system, shown in Fig. \ref{fig:level_diagram}.  An electromagnetic field couples two low-energy (qubit) states off-resonantly through a higher excited state, and the latter can be adiabatically eliminated in the process \cite{Lambda}. For example, in Fig. \ref{fig:level_diagram}, two laser beams with frequencies $\nu_{a}^{L}$ and $\nu_{b}^{L}$ couple the states $|a\rangle$ and $|b\rangle$ off-resonantly to the excited state $|e\rangle$, which is separated from the low energy states by frequencies $\nu_{a}$ and $\nu_{b}$ respectively. Stimulated Raman transitions will then be driven between the states $|a\rangle$ and $|b\rangle$ when the frequency difference between the beams $\nu_{a}^{L}-\nu_{b}^{L}$ is tuned close to the qubit resonance frequency $\nu_{ab}=\nu_{a}-\nu_{b}$. 

One convenient way to generate the requisite frequency difference is to split the light from a laser source into two paths, frequency shift the beam in one of the paths, and recombine them at the position of the qubit.  This eliminates sensitivity to common-mode fluctuations in the frequency of the laser \cite{Raman}. If the transition frequency $\nu_{ab}\lesssim 1$~GHz, an AOM may be used to generate the frequency shift.  For larger shifts up to $\sim10$ GHz, an electro-optic modulator can be used, although the resulting optical beat notes can be suppressed due to the phase-shifted pattern of the frequency-modulation sidebands \cite{PattyOptLett}.  
Mode-locked  lasers produce amplitude-modulated (AM) sidebands that do not suffer from this phase problem, and provide the best solution for generating the sidebands for this purpose. Bandwidths of mode-locked lasers are typically large enough to address frequency splittings $\nu_{ab}<100$ THz, and combined with modest frequency shifts from an AOM, they can be used to reach arbitrary frequency splittings within this bandwidth.  

\begin{figure}
\begin{center}
\includegraphics[width=0.45\textwidth]{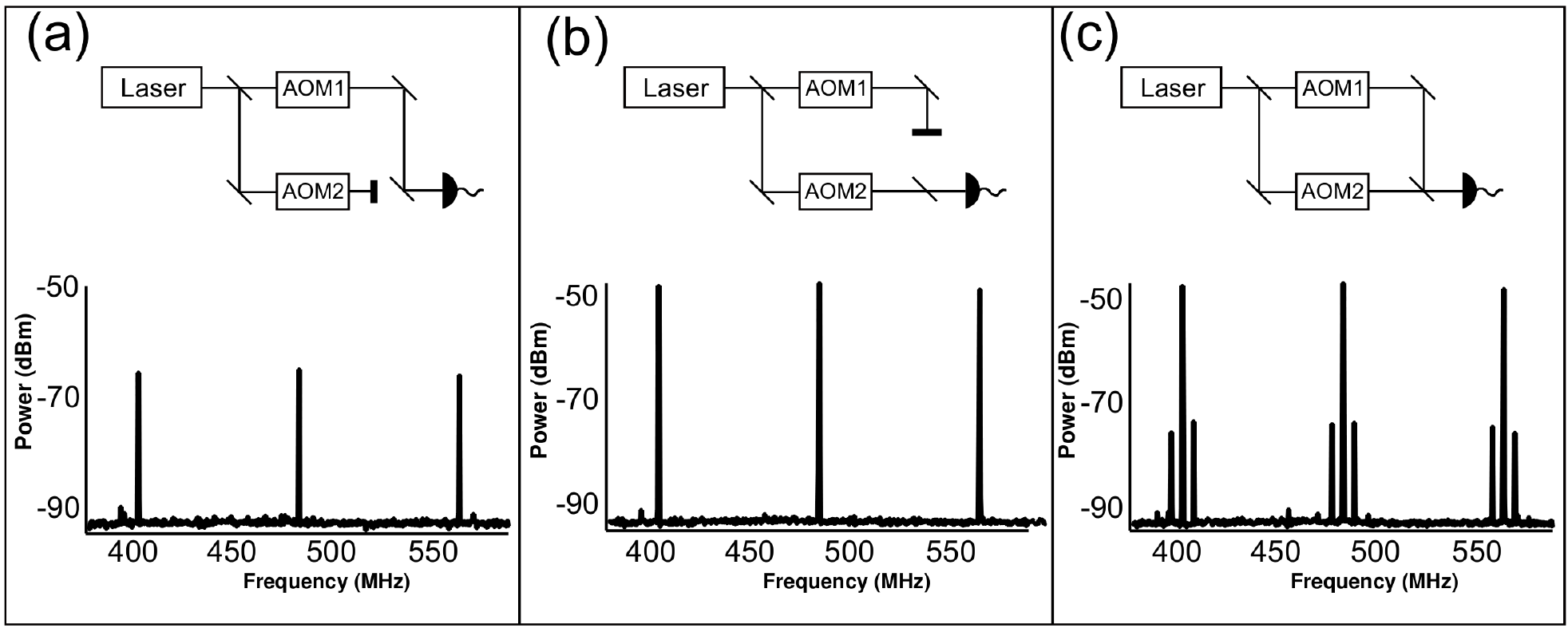}
\end{center}
\caption{Measured rf comb teeth from a mode-locked laser pulse train striking a fast photodiode. The light is split with a beamsplitter into two arms having separate AOMs, and recombined at the second beam-splitter with the photodiode behind one of the ports. (a) Partial spectrum with the beam from AOM2 blocked, showing lines spaced by the laser repetition frequency 
$\rep \sim 80$ MHz. (b) Spectrum with the beam from AOM1 blocked. (c) Spectrum with both beams, having sidebands at $\pm\dn$ due to interference of the beams. When a qubit replaces the beamsplitter, these rf beat notes can drive transitions between the qubit states.} 
\label{fig:combteeth}
\end{figure}

In order to tune the frequency of the fields that drive transitions in the qubit, the beam from a mode-locked laser is split into two paths with frequency shifts $\nu_{M1}$ and $\nu_{M2}$ impressed in the respective arms, and then recombined at the qubit.  The relative path lengths of the arms must be set to ensure simultaneous pulse overlap on the qubit, but they need not be long-term stable at optical wavelength scales.
Figure \ref{fig:combteeth} shows the resulting rf spectrum of the beams as measured by a fast photodiode,  
displaying the rf comb teeth separated by the laser repetition rate $\rep$ for each individual beam and then 
with additional beat notes separated by $\rep\pm\dn$ for the combined beams, where $\dn=|\nu_{M1}-\nu_{M2}|$.
We control the position of these extra sidebands by adjusting the relative shift of the paths $\dn$ to bring the beat note between a comb tooth and a sideband on resonance with an atomic transition: $\nu_{sb}=n\rep-\dn$ is set to $\nu_{ab}$, where $n$ is some integer and $\nu_{sb}$ is within the bandwidth of the frequency comb.
Note for $\Lambda-$transitions as in Fig. \ref{fig:level_diagram}, the laser carrier-envelope phase need not be stabilized \cite{Ye2007}, although this too can be accomplished in a similar manner as reported here \cite{Koke2010}.

\begin{figure}
\begin{center}
\includegraphics[width=0.35\textwidth]{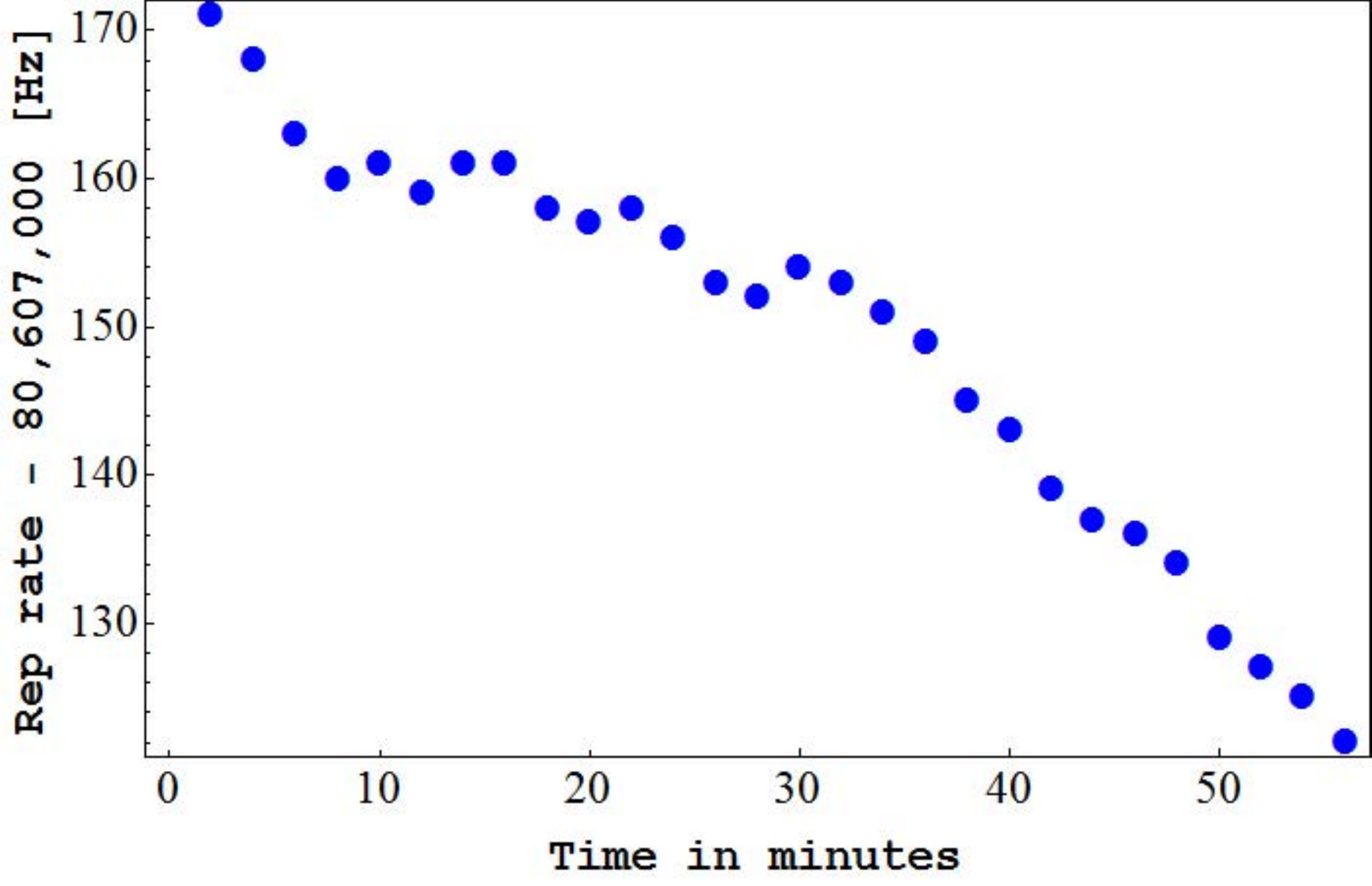}
\end{center}
\caption{Repetition rate of a free-running Nd:YVO$_4$ mode-locked laser at 355~nm (Vanguard from Spectra Physics), measured with a digital frequency counter in regular two-minute intervals. The drift of $\sim1$ Hz/m appears to be correlated with ambient temperature drifts. Raman transitions between \Yb qubit states use the $157^{\rm{th}}$ comb tooth, which therefore drifts by 
$\sim157$~Hz/min.} 
\label{fig:Rep_Rate_Drift}
\end{figure}

Driving the qubit requires optical coherence between the various comb teeth over the ms or faster time scale of the qubit manipulation, which is easy to achieve for typical transform-limited mode-locked laser sources 
\cite{HayesComb}.  However, as the laser repetition rate drifts [$\rep=\rep(t)$] due to
thermal or other mechanical strains in the laser cavity length (Fig. \ref{fig:Rep_Rate_Drift}), the applied sideband frequency $\nu_{sb}$ and hence the qubit drive will also drift from resonance. Stabilizing the repetition rate by directly feeding back an error signal to the laser cavity length may be difficult when the laser cavity is not easily accessible, as in the case of a hermetically sealed or fiber laser.  Furthermore, such a lock will have limited bandwidth, as the acquisition time for measuring the laser repetition rate and the delay in modulating the laser cavity length with a mechanical transducer may be longer than the characteristic time over which the laser cavity fluctuates. As an example, the hyperfine transition in the electronic ground state of \Yb (at 12.642819 GHz)  is near the $n=157^{\rm{th}}$ comb tooth of a mode-locked frequency tripled Nd:YVO$_4$ laser at $\rep \sim 80.6$ MHz. To stabilize this high order sideband within a fraction of 1 KHz, we must measure $\rep(t)$ to better than a few Hz, which corresponds to an integration time of longer than 1 s. We have found that this type of slow lock is insufficient for achieving high-fidelity qubit transitions with the frequency comb. 

\begin{figure}
\begin{center}
\includegraphics[width=0.4\textwidth]{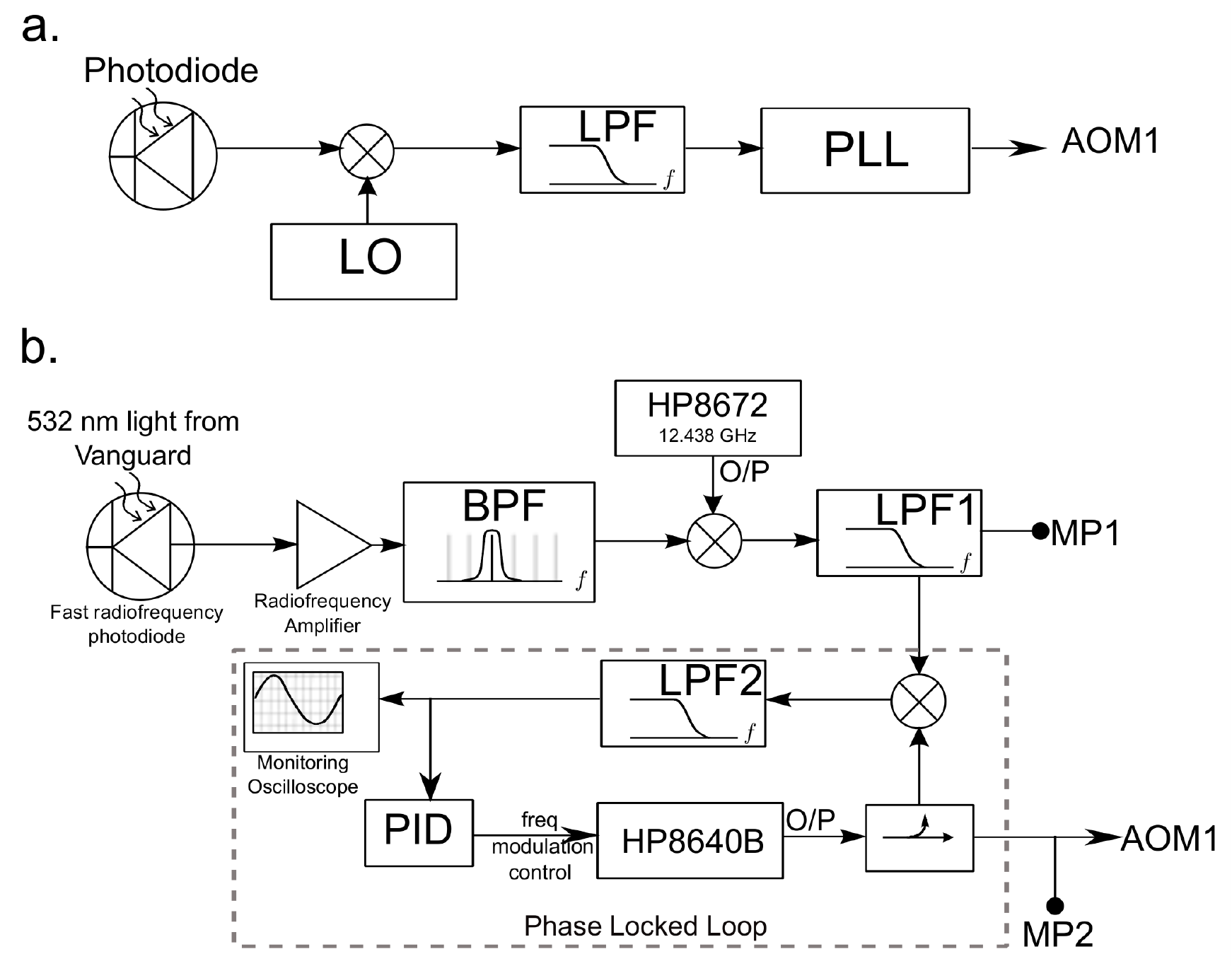}
\end{center}
\caption{Schematics of the beat note frequency lock. (a) The output of a fast photodiode is mixed with a local oscillator (LO) signal, which is sent to a Phase Locked Loop (PLL) after rejecting the high frequency beat note by using a low pass filter (LPF). The output of the PLL drives the AOM. (b) Details of the repetition rate lock circuit used in our experiment. The second harmonic light (at 532 nm) from a mode-locked YAG laser (Vanguard from Spectra Physics, repetition rate $\rep\sim 80.6$ MHz) is incident on a fast photodiode, which generates an rf comb with comb teeth at frequencies $m\rep$ ($m$ is a positive integer). This signal is amplified and passed though a bandpass filter (BPF),  which transmits the $n=157^{\rm{th}}$ comb tooth at $n\nu_{rep}\sim 12.655$ GHz. This is then
mixed with an rf signal at $\nu_{LO}=12.438$ GHz generated by an HP8672A oven-stabilized synthesizer, and the lower frequency beat note (at $\sim 217$ MHz) is sent to the PLL, where an HP8640B function generator is frequency modulated to produce a signal that is phase locked with the beat note. The bandwidth of the output signal depends on the bandwidth of the low pass filter LPF2 used in the PLL. The frequency spectra of the signals at monitoring points MP1 and MP2 are shown in Fig. \ref{fig:error_signal}.} 
\label{fig:lock_schematics}
\end{figure}

Instead, we continuously monitor the laser repetition rate by beating the photodiode signal at $n\rep(t)$ with a frequency-stabilized local oscillator at frequency $\nu_{LO}$, as shown in fig. \ref{fig:lock_schematics}(a). The beat note is sent through a filter which passes the lower frequency component of the beat signal at $n\rep(t)-\nu_{LO}$ (we assume that $n\rep(t)>\nu_{LO}$).  We can then electronically amplify this error signal and use it to drive one of the AOMs at frequency $\nu_{M1}(t)= n\rep(t)-\nu_{LO}$, so that fluctuations in the repetition rate are canceled and the sideband frequency appears at $\nu_{sb}=n\rep(t)-\dn=\nu_{LO} +\nu_{M2}$, which is independent of time.  Note that the fluctuations do not cancel for any other comb tooth or its sidebands in general. 

\begin{figure}
\begin{center}
\includegraphics[width=0.39\textwidth]{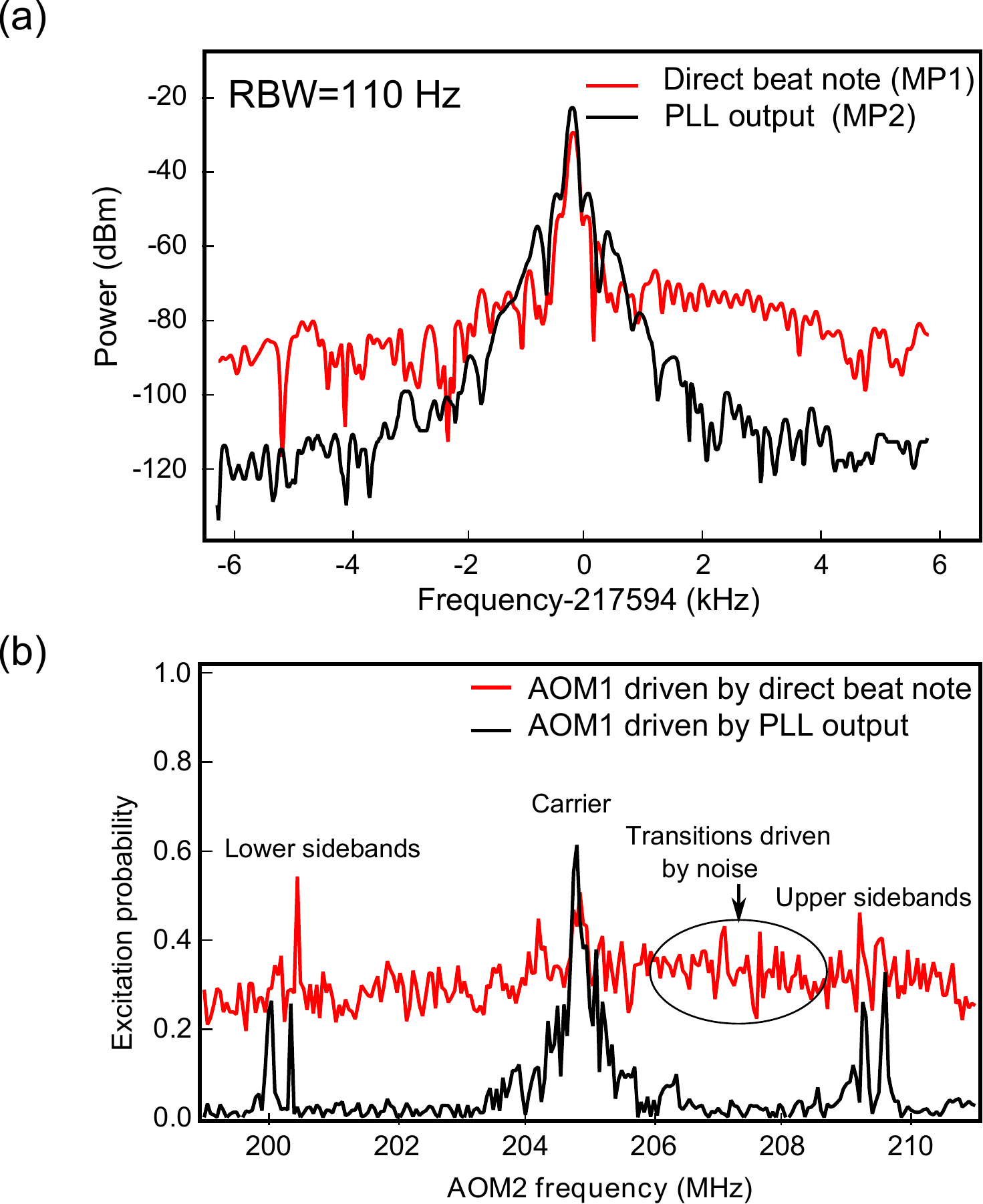}
\end{center}
\caption{(a) Frequency spectra of the error beat note that drives a modulator in one of the arms (AOM1) at points MP1 and MP2 in Fig. \ref{fig:lock_schematics}, with resolution bandwidth (RBW) of 110 Hz. The PLL gets rid of the (white) noise outside the bandwidth of the low pass filter, resulting in a fractional noise figure of $\alpha \sim -120$ dB/Hz. (b) Probability of Raman excitation of a single trapped \Yb ion qubit vs AOM2 frequency with and without the PLL. Here the system is initialized in the state 
$|a\rangle$. If the error signal at point MP1 is used to directly drive modulator AOM1, the noise excites unwanted transitions at all AOM2 frequencies, as seen in the constant background in the Raman spectrum. The output of the PLL does not have this noise beyond the bandwidth of the low pass filter used, and hence the Raman frequency spectrum is cleaner. Here we show the `carrier' transition between the hyperfine \Yb `clock' states appearing at a frequency of 
$\sim205$ MHz, and trapped ion vibrational upper and lower sidebands near $\sim200$ MHz and $\sim210$ MHz.}
\label{fig:error_signal}
\end{figure}

In practice, this error signal may have significant broadband noise from the amplifier circuit, causing fidelity errors in the qubit operation.  If the fractional spectral noise density of optical power in the error signal beam is $\alpha$~[dB/Hz] and assumed to be flat, the broadband Rabi frequency noise density will be $\Omega_0 \sqrt{\alpha}$ since the nominal Rabi frequency $\Omega_0$ is a product of the optical fields of the two beams and only one arm has this added noise.  The resulting probability of error in state evolution will scale as $\epsilon \sim (\pi^2/2)\alpha\Omega_0^2 T$, where $T$ is the total evolution time of the quantum process. 
For a Rabi frequency $\Omega_0=600$ kHz and an evolution time of $T=1$ ms, in order to keep $\epsilon<1\%$ we must have a fractional noise level of $\alpha<-115$ dB/Hz, which is difficult with realistic amplifiers.

However, it is possible to attain such noise floors by feeding the beat note error signal into a phase locked loop (PLL) configuration with a separate low-noise rf oscillator, and using its output to drive the modulator (AOM1). Beyond the bandwidth of the low-pass filter used in the PLL, the noise profile then takes the characteristics of the oscillator. Fig. \ref{fig:error_signal}(a) shows the measured spectrum of the error signal that drives the modulator AOM1 with and without the PLL, with a reduction of the noise floor by $\sim30$ dB with the PLL method. In fig. \ref{fig:error_signal}(b) we show the observed Raman spectrum of a single trapped \Yb ion taken by scanning the frequency of the beat note through AOM2, when the qubit is excited at a pulse duration of $40 \mu s$. The white noise present when directly driving AOM1 with the error signal produces unwanted Raman
transitions for a range of frequencies of AOM2, thus providing a non-zero background in the observed spectrum, as shown in the red trace. On the other hand, when we drive AOM1 with the output of the PLL, this background vanishes, as seen in the black trace. 

To characterize the stability of the beat note frequency, we compare it to the \Yb `clock' hyperfine qubit by Ramsey interferometric measurements. With the lock engaged we measure a coherence time of order $1$ s, consistent with known external sources of noise such as magnetic field noise \cite{OlmschenkYb}.  With the repetition rate lock disengaged, the coherence time drops to $\sim 3 $ms. 
Because the lock continuously tracks the difference frequency between a laser beat note and a microwave standard,  the lock speed is limited only by electronic delays and can have a bandwidth as high as $\sim~$100 MHz, while also tracking long term drifts without running out of range. 

\vspace{-12pt}
\noindent
\section*{ACKNOWLEDGEMENTS}
\vspace{-6pt}
This work is supported by the DARPA Optical Lattice Emulator Program, the IARPA MQCO Program under ARO contract, the ARO MURI Program on Hybrid Quantum Circuits, and the NSF Physics Frontier Center at JQI.

\vspace{-24pt}
%

\end{document}